\documentclass[twocolumn,showpacs,preprintnumbers,amsmath,amssymb]{revtex4}
\usepackage{stmaryrd}
\usepackage{txfonts}
\usepackage{amssymb}
\usepackage{graphicx}% Include figure files
\usepackage{dcolumn}% Align table columns on decimal point
\usepackage{bm}% bold math

\begin{document}
\title{ Phase-Sensitive measurements on the corner junction of  iron-based superconductor BaFe$_{1.8}$Co$_{0.2}$As$_{2}$}
\author{Yu-Rong Zhou$^1$, Yan-Rong Li$^{1,2}$, Jun-Wei Zuo$^1$, Rui-Yuan Liu$^1$, Shao-Kui  Su$^1$,\\ G. F. Chen$^1$, J. L. Lu$^1$, N. L. Wang$^1$, and Yun-Ping Wang$^1$ }
\email{ypwang@aphy.iphy.ac.cn} \affiliation{$^1$Beijing National
Laboratory for condensed matter Physics, Institute of Physics,
Chinese Academy of Sciences, P.O.Box 603, Beijing 100190,
People$^,$s Republic of China\\} \affiliation{$^2$Department of
Physics, University of Science and
technology of China, Hefei 230026, People$^,$s Republic of China\\}%

\date{\today}
\begin{abstract}
We have made a phase-sensitive measurement on the corner junction
of the iron-based superconductor BaFe$_{1.8}$Co$_{0.2}$As$_{2}$,
and observed the typical Fraunhofer-like diffraction pattern. The
result suggests that there is no phase shift between the {\it a-c}
face and {\it b-c} face of a crystal, which indicates that the
superconducting wavefunction of the iron based superconductor is
different from that of a cuprate superconductor.
\end{abstract}


\maketitle

The iron-based high T$_{c}$ superconductors discovered several
months ago have become the focus of interest in condensed-matter
physics because they have displayed the high T$_{c}$ up to 55K so
far \cite{Ren} and included a magnetic element in the crystalline
structure. The similarities between the iron-based and cuprate
superconductors, i.e. the layered crystal structure, the
approximate 2D conduction layer \cite{Kamihara} \cite{Rotter},
closeness to a long range antiferromagnetic order \cite{PCDai},
all suggest that the iron-based superconductors may have the same
superconducting mechanism as the cuprate superconductor's. Recent
 heat capacity measurement \cite{Kohama1} and
photoemission spectroscopy \cite{Kohama2} measurements seem to
favor
 this opinion. However, some other recent
experiments such as ARPES \cite{ZXShen, XJZhou, HDing}, infrared
spectrum \cite{GLi}, etc., would rather support that the
iron-based superconductors act more like a conventional
superconductor with regard to pairing behavior. Meanwhile, there
are two different kinds of theories in the effort of trying to
disclose the underlying superconducting mechanism: one is based on
the strong-coupling approach \cite{Haule, ZPYin, CCao}, which
emphasizes on-site correlations applicable to the high T$_{c}$
cuprate superconductors; the another is based on the weak-coupling
approach \cite{MHDu, SCZhang, Mazin}, which emphasizes
itinerant-electron physics. The debates indicate that there is the
need of much more work to do to determine the superconducting
wavefunction as well as its underlying mechanism in the iron-based
superconductor, among which the phase-sensitive experiment is
obviously one of the most important works drawing much attention.

The first phase-sensitive experiment on cuprate superconductor was
reported in 1993 \cite{VanHarlingenPRL93}. Since then,
phase-sensitive experiments based on different configurations
especially the corner junction \cite{VanHarlingenPRL95,
VanHarlingenRMP, TsueiPRL94, TsueiPRL, TsueiRMP, Ariando} have
played an important role in studying the high T$_{c}$ cuprate
superconductors, and it has been regarded as the most direct and
key tool in studying some intrinsic properties of the
superconducting wavefunction. Till now, it is still the only one
tool to directly detect the superconducting phase.

For an ideal corner junction based on a conventional
superconductor, its critical current as a function of magnetic
field is represented in the following form
\cite{VanHarlingenRMP}(shown in Fig.1a):
\begin{displaymath}
I_{c}=I_{0}\left|\frac{sin(\pi\Phi/\Phi_{0})}{(\pi\Phi/\Phi_{0})}\right|,
\end{displaymath}
where $\Phi$=\emph{Blt} is the total magnetic flux threading it,
\emph{l} is the length of the corner junction, \emph{t} is
magnetic barrier thickness, $\Phi_{0}$ is the flux quantum. It
reaches a peak at zero magnetic field.
\begin{figure}[hb]
\scalebox{0.55}{\includegraphics[bb=20 40 30cm 16cm]{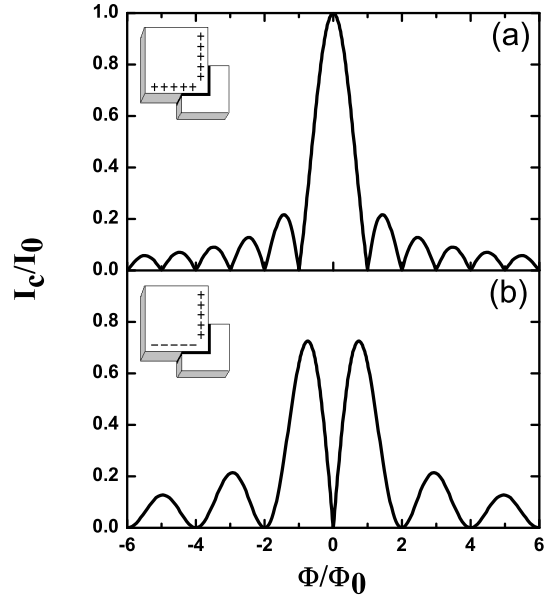}}
\caption{The schematic map of the diffraction pattern on a corner
junction: (a)  with zero phase shift; (b) with $\pi$ phase shift.}
\end{figure}

For a corner junction based on cuprate superconductor, its
critical current is represented in the different form
\cite{VanHarlingenRMP} (shown in Fig.1b):
\begin{displaymath}
I_{c}=I_{0}\left|\frac{sin^2(\pi\Phi/2\Phi_{0})}{(\pi\Phi/2\Phi_{0})}\right|.
\end{displaymath}
Obviously, at zero applied magnetic field, there is a minimum in
the curve of critical current,  because the $\pi$ phase difference
of the superconducting wavefunction between the two faces of the
crystal corner leads to a destructive interference of
superconducting current.

Therefore, the diffraction pattern of the critical current of
corner junction could be used as typical and direct evidence,
which indicates whether or not the wavefunction of a
superconductor is like that of the cuprate superconductor. In this
letter, we present the phase-sensitive measurement on the corner
junction of the iron-based superconductor
BaFe$_{1.8}$Co$_{0.2}$As$_{2}$. To our knowledge, this is the
first phase-sensitive experiment on the iron-based
superconductors.

\begin{figure}[hb]
\scalebox{0.35}{\includegraphics[bb=60 20 25cm 20cm]{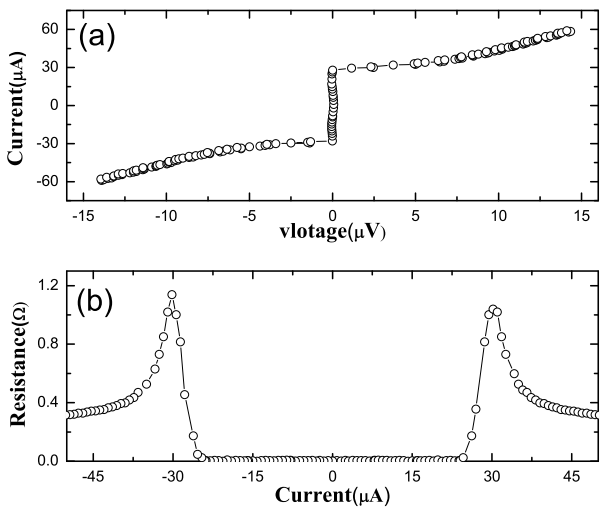}}
\caption{ The electrical characters of the corner junction at
1.8K: (a)Current-Voltage curve (b)Dynamic resistance vs voltage.}
\end{figure}

We fabricated single crystal BaFe$_{1.8}$Co$_{0.2}$As$_{2}$ into
corner junctions in the way described in Ref
\cite{VanHarlingenPRL93}. Our single crystals with T$_C$=22K was
obtained by flux-melt technique similar as described in
\cite{GFChen}. All of the samples are cleaved into small, thin
sheets with the typical thicknesses of $20\sim40\mu$m; and all the
faces used for corner junction are cleavage planes, smooth and
flat. After masking the sample (leaving the corner we need
uncovered which is even at the both faces), we sputter about 40nm
Au on the sample, then continue to sputter 300nm Pb over the Au
layer. The typical lengths of both sides of the corner junctions
in our experiment are 100$\sim$200$\muup$m, and the geometric
asymmetry of the corner junctions are less than 15$\%$(According
to Ref \cite{VanHarlingenPRL95}, the small asymmetry will not
affect the final conclusion).  Critical current of our junctions
used for measurement at 2 K is 20$\mu$A $\sim$ 3mA which is
feasible to measure at low temperature. We manufactured two
superconductor cans with inner layer of Pb and outer layer of Nb
in order to make sure of the good shielding effect. Measurement
was taken in the temperature range of 1.8K$\sim$ 4.2K which is far
below the transition temperatures of Nb and Pb.

The electrical characters of our corner junction is shown in
Fig.2, the I-V curve(Fig.2a) exhibits a typical resistively
shunted current-voltage character, which should be expected from
superconductor-normal metal-superconductor (SNS) junction with a
high quality. The very sharp transition in the dynamic resistance
curve (Fig.2b) makes it feasible to detect the critical current
precisely.

\begin{figure}[hb]
\scalebox{0.42}{\includegraphics[bb=80 50 30cm 19cm]{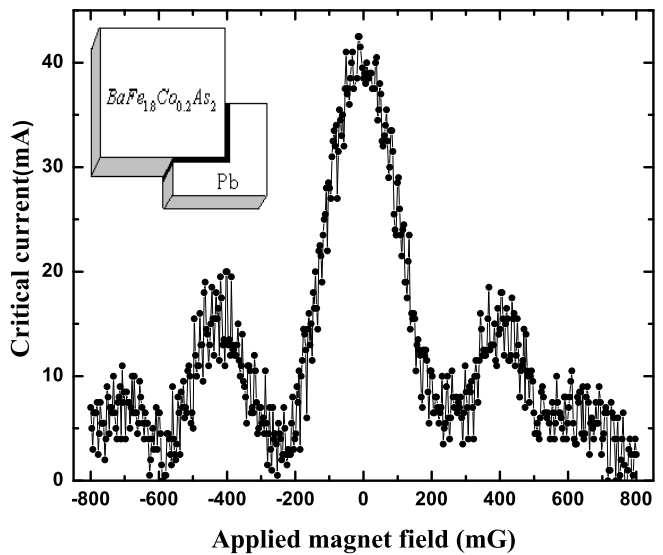}}
\caption{The Fraunhofer diffraction pattern of the critical
current as a function of magnetic field taken at 1.8K. Magnetic
field is applied by a self-made NbTi superconducting coil, the
maximum scanning range of magnetic field is -800mG$\sim$ 800mG,
limited by the Joule heat in contacts.}
\end{figure}

Magnetic field modulation of the critical current is shown in
Fig.3. It displays a typical, symmetric Fraunhofer diffraction
pattern, completely different from that of the corner junction of
cuprate superconductors, which shows a minimum instead of a
maximum at zero field. The diffraction pattern of our edge
junction is the same as that of our corner junction. Symmetric
Fraunhofer diffraction pattern of the corner junction indicates
that there is no phase shift between the {\it a-c} face and {\it
b-c} face of the corner. The possibility of flux trapping in the
corner could be  ruled out \cite{VanHarlingenPRL95}, because this
diffraction pattern can be repeatedly demonstrated in different
samples and in the same sample during several different thermal
cyclings between the measurement temperature and 25K.

It should be mentioned hereby that, there has been a common belief
of which the $\pi$ phase shift is a direct evidence for d-wave
pairing symmetry, and on the other hand, the zero phase shift is
that for s-wave pairing symmetry \cite{VanHarlingenRMP}
\cite{TsueiRMP}. Therefore, the result that we have reported here
seems to support
 s-wave pairing symmetry. However, some other points
of view \cite{YPWang}  is questioning about the ground theory
\cite{Larkin} \cite{Rice} of the above belief, and since the aim
of this letter is to report the experimental result, we will leave
the theoretical debate open for further research.

In summary, we did phase-sensitive experiment in the iron-based
superconductor BaFe$_{1.8}$Co$_{0.2}$As$_{2}$, the typical
Fraunhofer diffraction pattern shows that the critical current is
maximum at zero magnetic field, which means there is no phase
shift between the {\it a-c} face and {\it b-c} face of the crystal
corner. This indicates that the superconducting wavefunction of
the iron based superconductor is definitely not like that of a
cuprate superconductor.

This work was supported by the 973 project of Ministry of Science
and Technology of China, the National Natural Science Foundation of
China, and the Knowledge Innovation Project of Chinese Academy of
Sciences.

\end{document}